\documentclass[letterpaper, 10 pt, conference]{ieeeconf}  
\usepackage{amsmath}
\usepackage{algorithm}
\usepackage[noend]{algpseudocode}
\usepackage{algorithm}
\usepackage{algpseudocode}
\usepackage{pifont}
\usepackage{amsfonts}
\usepackage{subcaption}
\usepackage{kantlipsum}       
\usepackage{mwe}  
\usepackage{float}                 
\usepackage{graphicx}
\usepackage{caption}
\usepackage{subcaption}
\usepackage{textcomp}
\usepackage[english]{babel}
\usepackage{multicol}
\usepackage[noadjust]{cite}
\usepackage{textcomp}
\usepackage{mathtools}

\usepackage{hyperref}
\hypersetup{
    colorlinks=true,
    linkcolor=blue,
    filecolor=magenta,      
    urlcolor=cyan,
}

\usepackage[noadjust]{cite}                                                        

\IEEEoverridecommandlockouts                              
\title{\LARGE \bf
Deep Reinforcement Learning with Enhanced Safety for Autonomous Highway Driving
}

\author{Ali Baheri, Subramanya Nageshrao, H. Eric Tseng, Ilya Kolmanovsky, Anouck Girard, and Dimitar Filev
\thanks{Ali Baheri is with the Department of Aerospace and Mechanical Engineering, West Virginia University, Morgantown, WV 26505, USA. {\tt\small ali.baheri@mail.wvu.edu}}
\thanks{Subramanya Nageshrao is with Ford Greenfield Labs, 3251 Hillview Ave, Palo Alto, CA 94304, USA. {\tt\small snageshr@ford.com}}
\thanks{Ilya Kolmanovsky and Anouck Girard are with the Department of Aerospace Engineering, University of Michigan, Ann Arbor, MI 48109, USA. [{\tt\small ilya,anouck]@umich.edu}}
\thanks{H. Eric Tseng and Dimitar Filev are with Ford Research and Innovation Center, 2101 Village Road, Dearborn, MI 48124, USA.  [{\tt\small htseng,dfilev]@ford.com}}
}

\begin{document}

\maketitle
\thispagestyle{empty}
\pagestyle{empty}

\begin{abstract}

In this paper, we present a safe deep reinforcement learning system for automated driving. The proposed framework leverages merits of both rule-based and learning-based approaches for safety assurance. Our safety system consists of two modules namely handcrafted safety and dynamically-learned safety. The handcrafted safety module is a heuristic safety rule based on common driving practice that ensure a minimum relative gap to a traffic vehicle. On the other hand, the dynamically-learned safety module is a data-driven safety rule that learns safety patterns from driving data. Specifically, the dynamically-leaned safety module incorporates a model lookahead beyond the immediate reward of reinforcement learning to predict safety longer into the future. If one of the future states leads to a near-miss or collision, then a negative reward will be assigned to the reward function to avoid collision and accelerate the learning process. We demonstrate the capability of the proposed framework in a simulation environment with varying traffic density. Our results show the superior capabilities of the policy enhanced with dynamically-learned safety module.

\end{abstract}

\section{INTRODUCTION}

Safe decision making in automated driving is among the foremost open problems due to highly evolving driving environments and the influence of surrounding road users. Recent successes in reinforcement learning (RL) have motivated many researchers to apply RL in planning and decision making in autonomous driving. However, the question is, can we trust to an RL agent to achieve a certain level of safety in safety-critical applications?

A significant amount of effort has been given to address RL safety. Authors in \cite{garcia2015comprehensive} provide a comprehensive review of existing approaches. To address RL safety, an extensive body of literature has focused on transforming the optimization criterion to preserve the agent\textquotesingle s safety or avoid undesirable outcomes. For instance, \cite{coraluppi1997mixed} incorporates a penalty due to variability of a given policy that may potentially increase the possibility of undesirable behaviors. Similarly, \cite{sato2001td} transforms the main objective into a linear combination of return and variance of the return. Furthermore, \cite{altman1993asymptotic} introduces the idea of the constrained Markov process where the goal is to maximize the expectation of the return while maintaining other hard constraints.

\begin{figure}[t]
    \centering
    \includegraphics[width=0.5\textwidth]{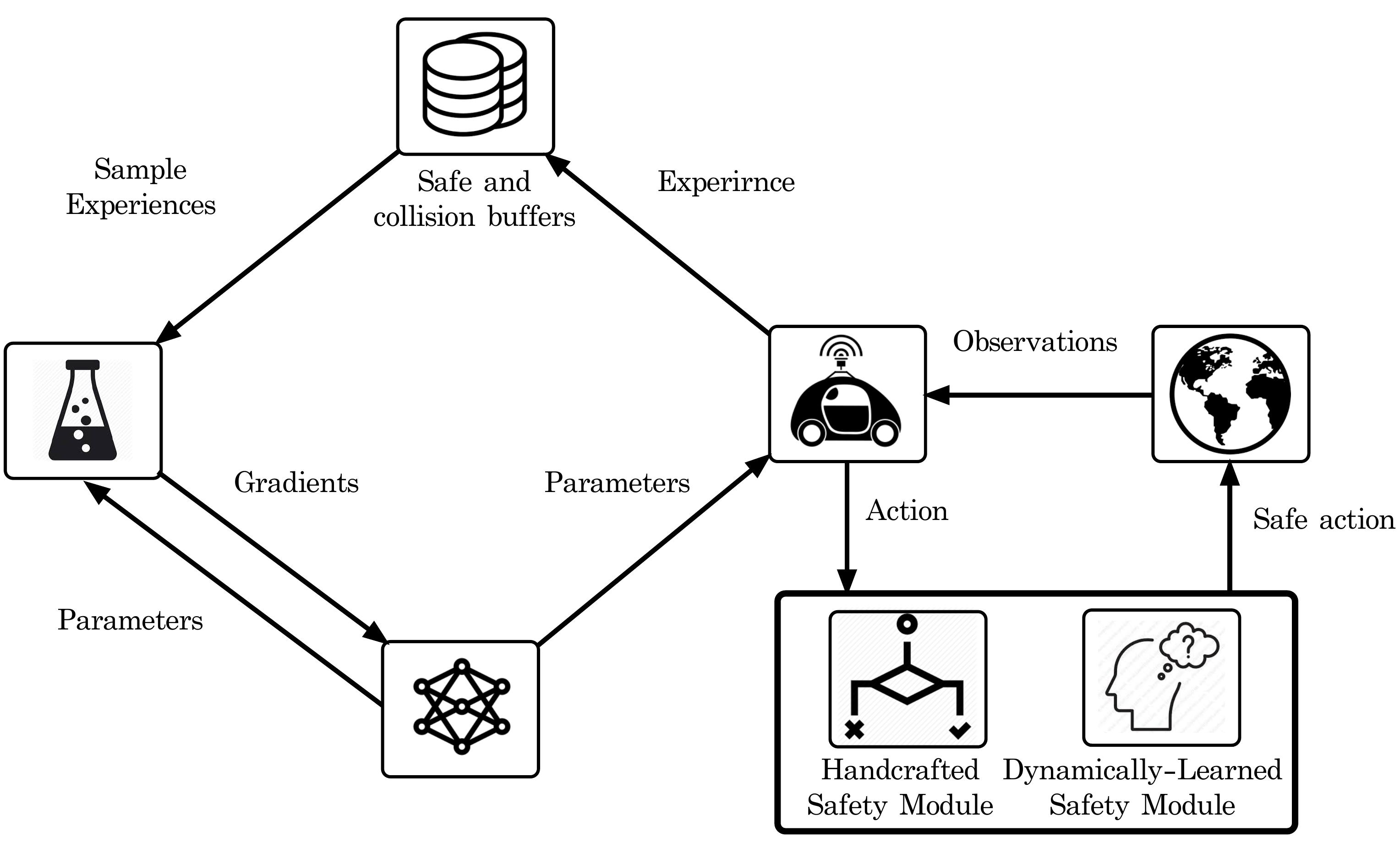}
    \caption{Imagine that a self-driving car is capable of predicting whether its future states are safe or one of them leads to a collision. Motivated by this scenario, we introduce a deep reinforcement framework enhanced with a learning-based safety component to achieve a more efficient level of safety for a self-driving car.}
    \label{fig:overall}
\end{figure}

Another line of research has focused on incorporating external knowledge to improve the safety of exploration. For instance, \cite{abbeel2010autonomous} derives a policy from a set of safe demonstrations. In a similar line of research, \cite{koppejan2011neuroevolutionary} provides initial information to RL for safe exploration. Some papers address safety RL through the use of risk-directed exploration \cite{law2005risk}. For example, \cite{gehring2013smart} introduces an algorithm to guide the exploration via a risk-measure to identify the probability of selecting different actions. The closest work to this research is \cite{lipton2016combating}, where inspired from the idea of intrinsically motivated RL \cite{chentanez2005intrinsically,barto2004intrinsically}, authors introduce the concept of intrinsic fear to guide exploration process to avoid undesirable behaviors.

To-date, the majority of research for safe automated driving has focused on rule-based approaches, inferred as handcrafted state machines. For instance, \cite{DBLP:journals/corr/abs-1708-06374} aims to formalize general requirements- called Responsibility, Sensitivity, Safety (RSS), that an autonomous vehicle must satisfy for safety assurance. However, in a highly dynamic and evolving environment, there is no guarantee that rule-based approaches prevent undesirable behaviors. Furthermore, rule-based approaches are not able to generalize to unknown situations. Comparatively, fewer studies have focused on the impact of incorporating external knowledge or models for safety assurance in the learning phase. 

This work is built on top of \cite{DBLP:journals/corr/abs-1904-00035}, where a complimentary safety module is incorporated into the learning phase. In contrast to the related papers, our framework benefits from the merits of both rule-based and learning-based safety approaches. Specifically, the additional safety component, called the \emph{dynamically-learned safety module}, incorporates a model lookahead into the learning phase of RL to predict safety longer into the future and to determine whether the future states lead to undesirable behaviors or not. If one of the future states leads to a collision, then a penalty will be assigned to the reward function to prevent collision and to reinforce to remember unsafe states. 

Furthermore, thanks to the nature of our state representation, the proposed framework takes into account the intentions of other road users into the decision making. We argue that providing a penalty for those predicted undesirable states accelerates the learning process. To learn safety rules in an evolving environment, we train a recurrent neural network (RNN) using historical driving data to temporally learn about the \emph{safety patterns} from the surrounding environment.

The main contributions of this paper can be summarized as follows:
\begin{itemize}
\item We propose an architecture for safe automated driving that takes into account the promising features of both handcrafted and dynamically-learned safety modules.

\item We argue that the dynamically-learned safety module not only predicts safety longer into the future, but also learns to predict the future motions of other road users and incorporates them into the decision making phase.

\item We illustrate the capability of the proposed framework in a simulation environment with varying traffic density and demonstrate the effectiveness of the dynamically-learned safety module. 

\end{itemize}
The remainder of this paper is structured as follows: In Section \ref{ref:problem statement}, we introduce the problem at hand and formalize the RL problem. Section \ref{sec:method} introduces our approach, including detailed illustrations of handcrafted and dynamically-learned safety modules. Section \ref{sec:results} provides detailed results, demonstrating improvement over the previous work.

\section{Problem Statement and System Architecture}
\label{ref:problem statement}

The objective considered in this paper is to design a learning framework capable of addressing safety concerns for an autonomous vehicle in a three-lane highway scenario. 

Fig. \ref{fig:overall} depicts our proposed architecture. We formalize the problem as a Markov decision process (MDP) where at each time-step $t$, the agent interacts with the environment, receives the state $s_t \in \mathcal{S}$, and performs an action $a_t \in \mathcal{A}$. As a result, the agent receives a reward $r_t \in \mathcal{R}$ and ends up in a new state $s_{t+1}$. The goal is to find a policy, $\pi$, that maps each state to an action with the goal of maximizing expected cumulative reward, $\sum_{k = 0}^{\infty}\gamma^{k}r_{t+k} $, where $0 < \gamma < 1$, is the discount factor \cite{sutton1998introduction}.

The optimal action-value function, $Q^{\ast}(s,a)$, obeys the following identity known as Bellman equation,

\begin{equation}
Q^{\ast}(s,a) =\underset{{s}'} {\mathbb{E}} \ [r+ \gamma \  \underset{{a}'}\max \  Q^{\ast}({s}\textquotesingle,{a}\textquotesingle)|(s,a)].
\end{equation}
For the small scale problems, $Q^{\ast}(s,a)$, is efficiently estimated using tabular methods. For large state space problems, however, a function approximator is utilized to approximate the optimal action-value function. Approximating the optimal action-value function using a neural network associated with a few other tricks to stabilize the overall performance build the foundation of double deep Q-network (DDQN) which serves as our decision making engine in this work \cite{mnih2015human,van2016deep}.

\begin{figure}[t]
    \centering
    \includegraphics[width=0.45\textwidth]{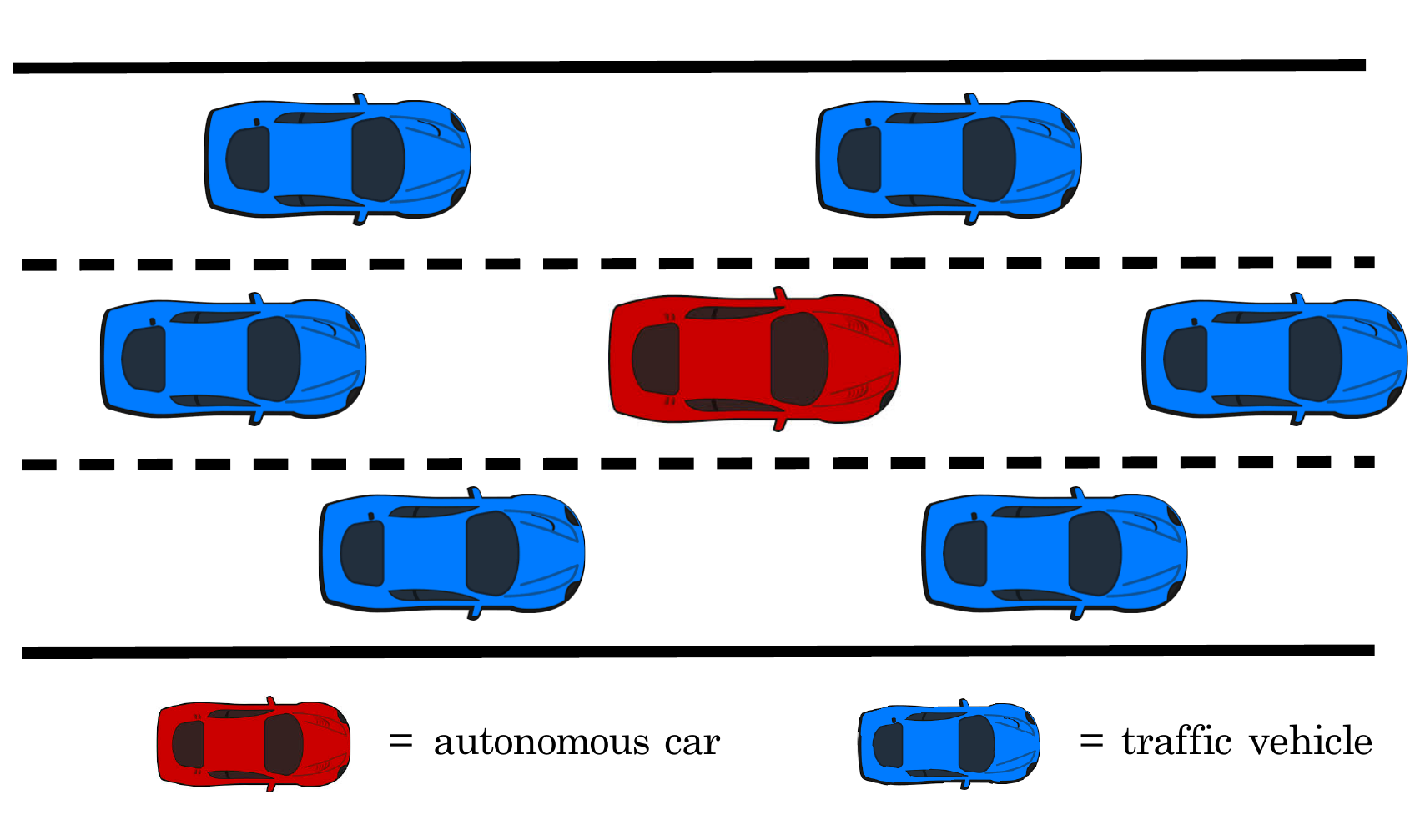}
    \caption{We study the problem of safe decision making for an autonomous car in a three-lane highway driving scenario where the autonomous agent can be surrounded by up to six traffic vehicles.}
    \label{fig:highway}
\end{figure}

\subsection{State space}

We assume a direct perception based approach to estimate the affordance for driving from \cite{chen2015deepdriving} for state representation. Fig. \ref{fig:highway} demonstrates the scenario considered in this paper where the autonomous car is surrounded by up to six traffic vehicles in a three-lane highway. A total of $18$ affordance indicators are used to spatiotemporally represent the information of the six nearest traffic vehicles. These variables include the relative distance and velocity, in longitudinal direction, to the nearest front/rear car in the right/center/left lane from the autonomous vehicle\textquotesingle s perspective. 

In addition to those indicators, we use longitudinal velocity and lateral position of the autonomous vehicle. In total, these $20$ affordance indicators represent a minimal yet sufficient state representation for the three-lane highway driving scenario studied in this work.

\subsection{Action space}

We consider four action choices along longitudinal direction, namely, maintain, accelerate, brake, and hard brake. For lateral direction we assume three action choices, one for lane keep, change lane to right, and change lane to left. These result in $8$ unique action choices. 

\subsection{Reward function}

The reward function is formulated as a function of (i) desired traveling speed subject to traffic condition, (ii) desired lane and lane offset subject to traffic condition, and (iii) relative distance to the preceding car based on relative velocity as follows:

\begin{equation}
r_v = e^{-\frac{(v_{e_x} - v_{\mathrm{des}})^2}{10}} - 1,
\end{equation}
\begin{equation}
r_y = e^{-\frac{(d_{e_y} - y_{\mathrm{des}})^2}{10}} - 1,
\end{equation}
\begin{equation}
  r_x=\begin{cases}
    e^{-\frac{(d_\mathrm{lead} - d_{\mathrm{safe}})^2}{10d_\mathrm{safe}}} - 1, & \text{if $d_\mathrm{lead}<d_\mathrm{safe}$},\\
    0, & \text{otherwise},
  \end{cases}
\end{equation}
where $v_{e_x}$, $d_{e_y}$, and $d_{\mathrm{lead}}$ are the autonomous agent\textquotesingle s velocity, lateral position, and the longitudinal distance to the lead vehicle, respectively. Similarly, ${v_\mathrm{des}}$, $y_{\mathrm{des}}$, and $d_{\mathrm{safe}}$ are the desired speed, lane position, and safe longitudinal distance to the lead traffic vehicle, respectively.

\subsection{Vehicle dynamics}

We model each vehicle using a computationally efficient point-mass model. For longitudinal equations of motion we use a discrete-time double integrator,

\begin{equation}
x(t+1) = x(t)  +v_x(t) \Delta t,
\end{equation}
\begin{equation}
v_x(t+1) = v_x(t)  +a_x(t) \Delta t,
\end{equation}
where $t$ is the time index, $\Delta t$ is the sampling time, and $x$ is the longitudinal position.  $v_x$ and $a_x$ are the longitudinal velocity and longitudinal acceleration of the vehicle, respectively. For the lateral motion, we assume a simple kinematic model,

\begin{equation}
y(t+1) = y(t)  +v_y(t) \Delta t.
\end{equation}
where $y$ is the lateral position of the car.



\section{Methodology: Safety Modules}
\label{sec:method}

\begin{figure*}[t]
    \centering
    \includegraphics[width=0.65\textwidth]{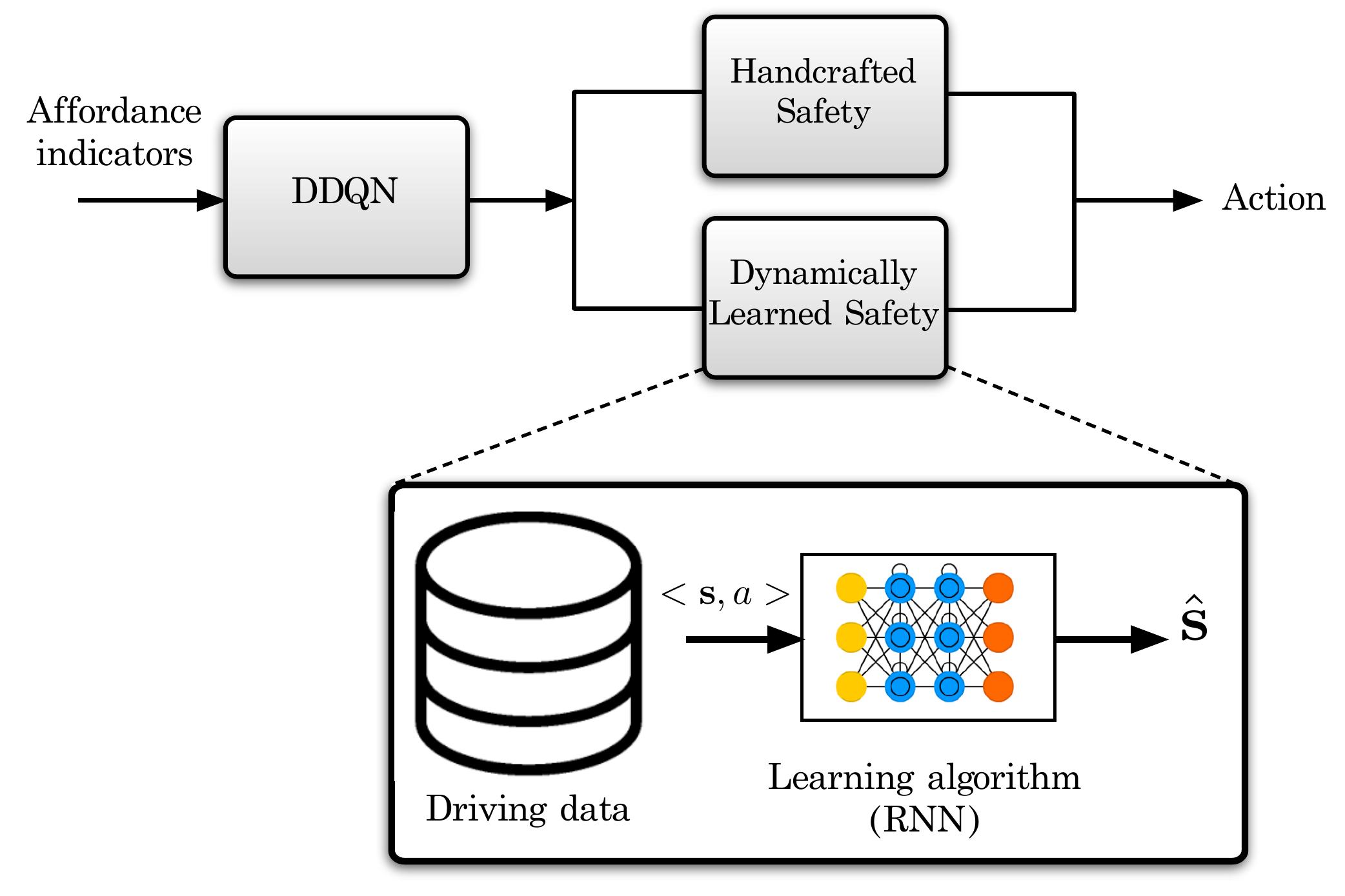}
    \caption{The proposed architecture for safe automated driving. Our algorithm leverages the merits of both rule-based and learning-based approaches. The rule-based module, called handcrafted safety, is based on common driving practice, while the learning-based module serves as the model lookahead to predict safety longer into the future. To predict future states, we train an RNN based on historical driving data whose inputs are a history of states and actions, and outputs are predicted future states for a given horizon. We assign a negative reward to the reward function, if one of the predicted future states violates the safety rules.}
    \label{fig:framework}
\end{figure*}

We address the problem of safe autonomous driving by introducing an additional safety module that aims to encode prior knowledge about the environment into the learning phase. The overall architecture of the proposed driving system is illustrated in Fig. \ref{fig:framework}. The system consists of two safety components, namely, the \emph{handcrafted safety module} and \emph{dynamically-learned safety module}. The first module is a heuristic safety rule based on common driving practice. On the other hand, the second module intends to predict safety longer into the future via an offline trained supervised model to guide the exploration process. We describe each of the two modules in detail in the next subsections.

\begin{algorithm*}
\caption{DDQN with enhanced safety module}
\begin{algorithmic}[1]
\State \textbf{Inputs:} Trained RNN, prediction horizon $k$
\State \textbf{Initialize:} Safe buffer, collision buffer, $Q$-network, and target $Q$-network
\While{not done}
\State Initialize cars and obtain affordance indicators $s$
\For {length of an episode or collision}
\State Perform $\epsilon$-greedy and select action $a_t$
 \If{action is not safe}
 \State Store $(s_t,a_t,\ast,-R_{handcraft})$ in collision buffer and replace by safe action
 \State Apply action, observe the next state, and obtain reward
 \EndIf
 \If{Static collision}     {\Comment{Handcrafted safety module}}
 \State Reward $\leftarrow$ $-R_{handcraft}$
 \State Store $(s_t,a_t,\ast,-R_{handcraft})$ in collision buffer
 \Else 
 \State Store $(s_t,a_t,s_{t+1}, r_{t+1})$ in safe buffer
 \EndIf
 \State Use RNN to predict $\hat{s}_{t+1}, \hat{s}_{t+2}, \dots , \hat{s}_{t+k}$  {\Comment{Dynamically-learned safety module}}
 \If{Dynamic collision for \emph{any} future (predicted) states}   
 \State Reward $\leftarrow$ $-R_{dynamic}$
 \State Store $(s_t,a_t,\hat{s}_{t+1}, -R_{dynamic})$ in collision buffer
 \EndIf
 
\State Sample random mini-batch $(s_\tau,a_\tau,s_{\tau+1},r_{\tau+1})$, $50\%$ from safe buffer and $50\%$ from collision buffer
\State Set $  y_{\tau}=\left\{
  \begin{array}{@{}ll@{}}
    r_{\tau+1} & \mbox{if sample is from collision buffer} \\
    r_{\tau+1} + \gamma \hat{Q}\Big(s_{\tau+1},\mathrm{argmax}_a \ Q(s_{\tau+1},a,\theta_\tau),\hat{\theta}_\tau\Big) & \mbox{if sample is from safe buffer}
  \end{array}\right.
$
\State Perform gradient descent on $\Big(y_\tau- Q(s_\tau,a_\tau,\theta_\tau)\Big)^2$ w.r.t $\theta$

\EndFor
\EndWhile
\end{algorithmic}
\label{alg:ddqn}
\end{algorithm*}
 
\subsection{Handcrafted safety module}
The handcrafted safety module is based on well-known traffic rules that ensure a minimum relative gap to a traffic vehicle based on its relative velocity,

\begin{equation}
d_{TV} - T_{\mathrm{min}} \times v_{TV} > {d_{TV}}_{\mathrm{min}},
\label{eq:static}
\end{equation}
where $d_{TV}$, $v_{TV}$ are the relative distance and relative velocity to a traffic vehicle, $T_\mathrm{min}$ is the minimum time to collision, ${d_{TV}}_{\mathrm{min}}$ is the minimum gap which must be ensured before executing the action choice. If this condition is not satisfied, an alternate safe action will be provided. Specifically,

\begin{equation}
  a_{\mathrm{safe}} =
  \begin{cases}
    \text{Hard brake} & \text{if $T_c \leq T_{hb}$}, \\
    \text{Brake} & \text{if $	T_{hb} < T_c \leq T_b$}, \\
    \text{Maintain} & \text{if $T_b < T_c$},
  \end{cases}
\end{equation}
where $a_{\mathrm{safe}}$ is the safe action and $T_c$ is the time-to-collision. $T_{hb}$ and $T_b$ represent the thresholds above which the decision made by the RL is considered to be
safe.

\subsection{Dynamically-learned safety module}

The handcrafted safety rules may not be sufficient to prevent collision since there is a wide variety in the behavior of surrounding vehicles. On the other hand, they may over-constrain an automated vehicle\textquotesingle s transient movement. The ultimate goal is to find complimentary safety rules on-the-fly to avoid accident prone states and accelerate the learning process. To that end, we train an RNN to predict whether future states lead to an accident within a pre-defined finite horizon. If one of the future states in a given horizon leads to an accident, then we conclude that the dynamically-learned safety rule is violated. In that case, we assign a negative reward to RL. In short, compared to the handcrafted safety rule, the dynamically-learned safety module learns to predict the future and thus accelerates the learning process.

\subsubsection{RNN Training}

To collect data for RNN training, we train an RL agent without dynamically-learned safety module. Once trained, we evaluate the policy and collect a long history of states and corresponding action that builds our driving data. To construct a predictor, we train an RNN whose inputs are the pair of states and action and its outputs is a vector of predicted future states. 

%
%
%

We summarize the DDQN with enhanced safety modules in Algorithm \ref{alg:ddqn}. The algorithm is initialized with two buffers, namely the safe and collision buffer, to store good and bad behaviors. At each time step, we check whether the handcrafted safety module is satisfied using Eq. \ref{eq:static}. If not, we store the danger state in the collision buffer and assign a large negative reward, $-R_\mathrm{handcraft}$, to the reward function. Otherwise, we store it in the safe buffer (line $9-13$). Next, we check whether the dynamically-learned safety module is satisfied by (i) predicting the future states in a finite horizon via the trained RNN and (ii) examining whether there exists any violation of safety rules (Eq. \ref{eq:static}). In case of violation, we store the next state in the collision buffer and assign a large negative reward, $-R_\mathrm{dynamic}$, to the reward function (line $14-17$). Then, to update the temporal difference target, we equally sample from both collision and safe buffers (line $18-19$). Finally, the model parameters are updated using a stochastic optimization algorithm.

\section{Results}
\label{sec:results}

We evaluate the effectiveness of the proposed framework in a simulation environment. Based on Algorithm \ref{alg:ddqn}, during the learning phase, the autonomous agent utilizes an $\epsilon$-greedy strategy to make decisions. Other traffic vehicles are controlled externally. Specifically, we use a combination of controllers from \cite{li2017game} and \cite{zhang2017finite} to control traffic vehicles. Furthermore, other system parameters such as maximum velocity are randomly chosen for traffic vehicles. The traffic vehicles are also allowed to randomly perform lane changes.

\begin{figure}[t]
    \centering
    \includegraphics[width=0.5\textwidth]{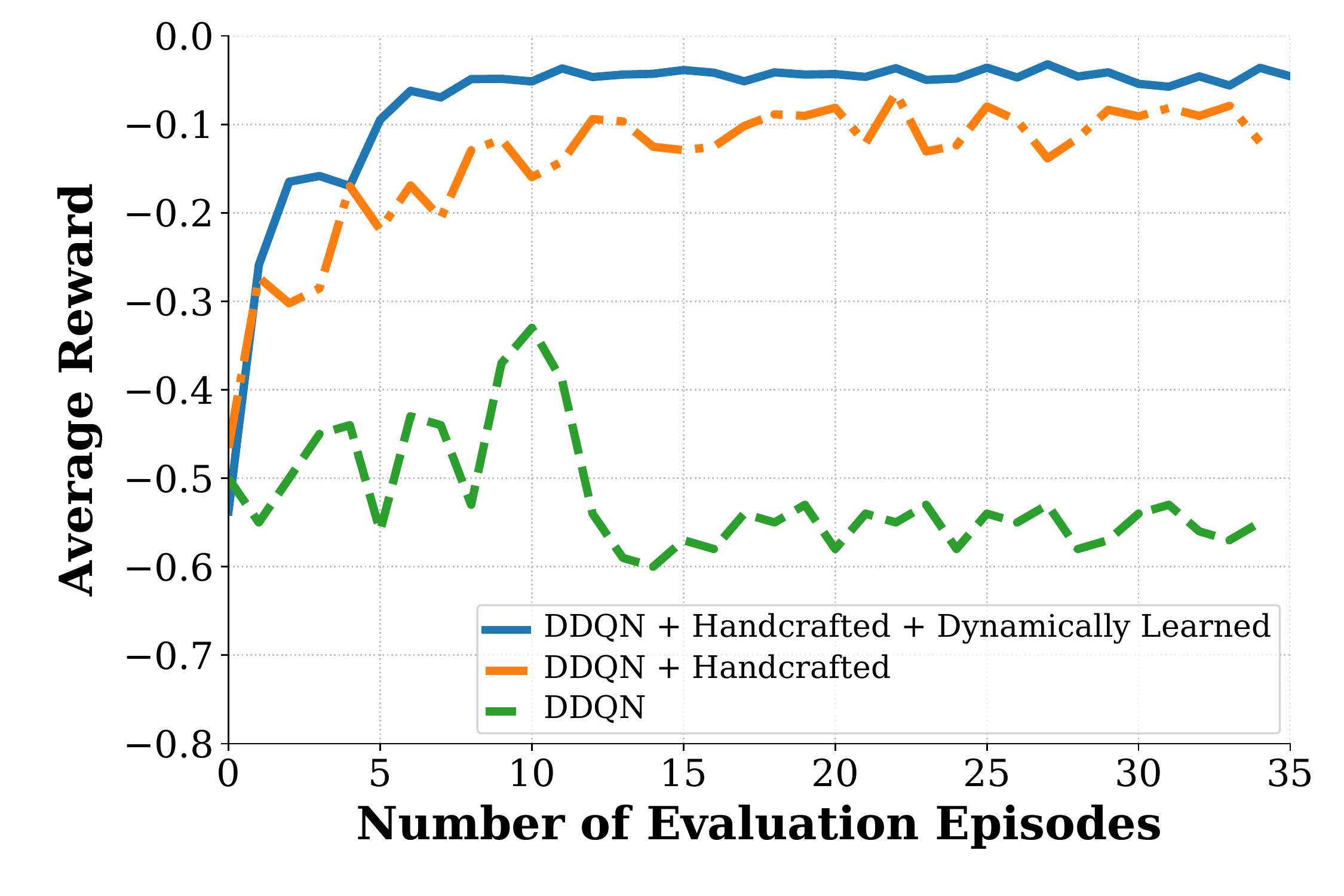}
    \caption{Learning curves for three policies during training. We train three policies (i) DDQN with both handcrafted and dynamically-learned safety modules (ii) DDQN with only handcrafted safety module and (iii) DDQN without any safety module, for a total of $3500$ episodes, evaluate them every $100^{\mathrm{th}}$ episode, and report the average cumulative reward.}
    \label{fig:learning_curve}
\end{figure}

We train our autonomous agent for a total of $3500$ episodes. Table \ref{table:train_params} presents the training parameters. Each episode is initialized with a randomly chosen number of traffic vehicles. Each episode terminates when the autonomous agent collides with a traffic vehicle, or when a time budget is exhausted. During learning, we partially evaluate the proposed architecture every $100^{\mathrm{th}}$ episode. Fig. \ref{fig:learning_curve} represents the cumulative reward during the training phase. One can conclude that the policy enhanced with dynamically-learned safety module outperforms the policy with only handcrafted safety module.

\begin{figure}[t]
    \centering
    \includegraphics[width=0.55\textwidth]{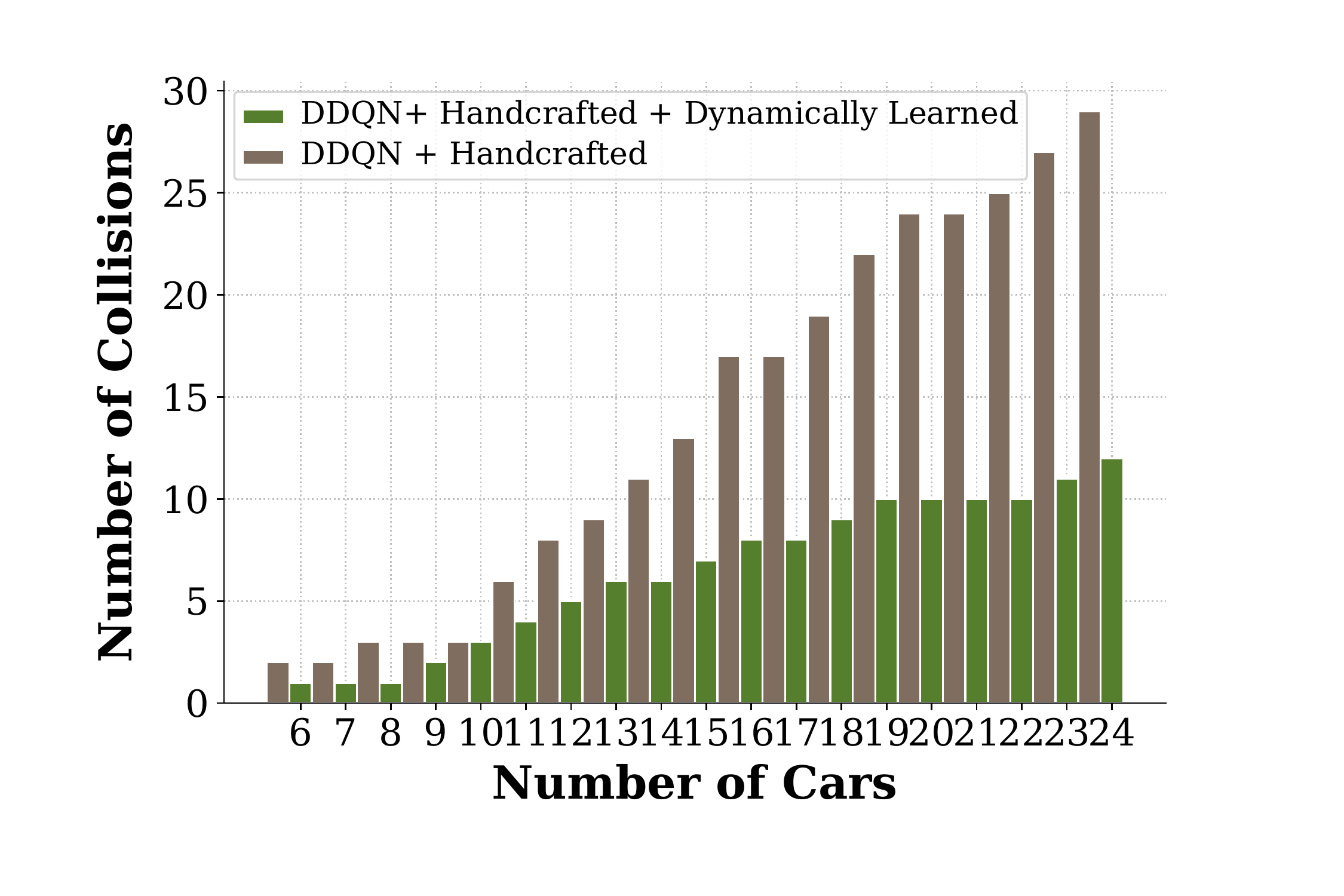}
    \caption{Number of collisions for different numbers of traffic vehicles after training. We evaluate two policies after training for $3000$ times. The brown bars report the number of collisions for the policy with only the handcrafted safety module, while the green bars represent the number of collisions for the policy with both handcrafted and dynamically-learned safety modules.}
    \label{fig:collision}
\end{figure}

Once trained, we freeze two policies and evaluate them with and without dynamically-learned safety module. Fig. \ref{fig:collision} represents the number of collisions for different numbers of traffic vehicles after training. We evaluate two policies for $3000$ times. The brown bars report the number of collisions for policy with only handcrafted safety module, whereas the green bars represent the number of collision for the policy with both handcrafted and the dynamically-learned safety modules. It can be seen that the policy enhanced with dynamically-learned safety module results in significantly fewer collisions. Fig. \ref{fig:collision} also shows that as the number of vehicles increases, the number of collisions increases, as expected.

\begin{table}[]
\centering
\caption{Training Parameters}
\begin{tabular}{cc}
\hline
\textbf{Number of input neurons}                                     & 20                       \\
\textbf{Number of output neurons}                                    & 8                        \\
\textbf{Number of hidden layers}                                     & 2                        \\
\multicolumn{1}{l}{\textbf{Number of neurons in each hidden layer}} & 100                      \\
\textbf{Learning rate}                                               & 1e-3                     \\
\textbf{Gamma}                                                       & 0.9                      \\
\textbf{Optimizer}                                                   & Adam                     \\
\textbf{Activation function}                                         & \multicolumn{1}{l}{leaky ReLU} \\
\textbf{Number of training episodes}                                 & 3500                     \\
\textbf{Number of time-step in each episode}                         & 200                      \\ \hline
\end{tabular}
\label{table:train_params}
\end{table}

\subsection{Discussion and future directions}

We argued that in a highly changing environment, in the presence of unexpected behaviors from other road users, the kinematic safety rules such as RSS are not sufficient to prevent undesirable situations. To that end, we proposed a learning-based safety module to learn existing safety patterns from historical driving data. We demonstrated that the additional safety module results in significantly fewer collisions. However, it should be recalled that in safety-critical applications such as self-driving cars even a single collision is not acceptable. Our ultimate goal is to further reduce the number of collisions. We plan to expand the current framework to address this issue. 

Thus far, we utilized an offline trained RNN to predict future states in the learning phase. What if we update our \emph{belief} about the predictor (i.e., RNN) as we learn more about the environment? We postulate that adjusting the RNN on-the-fly not only leads to fewer collisions, but also leads to an \emph{adaptive} decision making framework which is highly required for an evolving environment. To achieve that goal, we plan to employ an \emph{iterative training process} such that the dynamically-safety rule be frequently updated as we learn more about the environment. In an iterative training process setting, we train the agent for a number of episodes, say $100$, evaluate the policy and collect data to \emph{re-train} the RNN based on new information until convergence. To adjust the RNN model, we leverage algorithms from the incremental learning community where new information is incremented to existing knowledge in lieu of learning from scratch. Furthermore, algorithms such as \cite{hausknecht2015deep} can be used to formulate the problem as a partially observable Markov decision process (POMDP) to find a better policy in a highly changing environment. 

The proposed algorithm also can be implemented on advanced driver-assistance systems where state/action data is utilized for imitation learning of driver\textquotesingle s policy.
Specifically, once the RNN mapping is learned, it can be uploaded as a complementary safety model to the RL algorithm.

\section{CONCLUSIONS}

In this paper, we proposed a deep reinforcement learning architecture for safe automated driving in a three-lane highway scenario. We argued that heuristic safety rules are susceptible to deal with unexpected behaviors particularly in a highly changing environment. To alleviate this issue, we proposed a learning-based mechanism to learn safety patterns from historical driving data. To achieve that goal, we trained an RNN (i) to predict future states within a finite horizon and (ii) to determine whether one of the future sates violates the safety rule. We evaluated our approach in a simulation environment with varying traffic density. We demonstrated that incorporating a learning-based safety module accelerates the learning process and results in significantly fewer collisions.

\section{Acknowledgment}

This work is supported by Ford Motor Company.

\bibliographystyle{unsrt}
\bibliography{Bo_Summer2016}

\end{document}